\documentclass[aps,nofootinbib]{revtex4}

\usepackage{eurosym}
\usepackage{amsfonts}
\usepackage{amsmath}
\usepackage{amssymb}
\usepackage{xcolor}
\usepackage{float}
\usepackage{graphicx}
\usepackage[utf8]{inputenc}
\usepackage{hyperref}
\usepackage{amssymb,epsf}
\usepackage{latexsym}
\usepackage{epstopdf}
\usepackage{epsfig}
\usepackage{subfigure}

\begin{document}
\title{\bf $Z_{2}$-Symmetric Thick Brane \\with LambertW Warp Function}
\author{
 S. H. Hendi$^{1,2,4}$\footnote{Electronic address: hendi@shirazu.ac.ir},
 N. Riazi$^{3}$\footnote{Electronic address: n\_riazi@sbu.ac.ir},
 S. N. Sajadi$^{1,2,3}$\footnote{Electronic address: naseh.sajadi@gmail.com}
 }
\affiliation{
 $^1$Department of Physics, School of Science, Shiraz University, Shiraz 71454, Iran \\
 $^2$Biruni Observatory, School of Science, Shiraz University, Shiraz 71454, Iran    \\
 $^3$Department of Physics, Shahid Beheshti University, G.C., Evin, Tehran 19839,  Iran \\
 $^4$Canadian Quantum Research Center 204-3002 32 Ave Vernon, BC V1T 2L7 Canada}

\begin{abstract}
In this work, we investigate a new warped five-dimensional,
$Z_{2}$-symmetric thick brane solution in the presence of a real
scalar field. We examine the different geometric aspects of the
model. We discuss the stability of the solution under
gravitational fluctuations and study both the graviton ground
state and the continuum of Kaluza-Klein modes to find a correction
to Newton's law. Then, we study the quantum version of the model
by deriving the Wheeler-DeWitt equation and looking for the
corresponding solution. Finally, the cosmology of the brane is
studied.
\end{abstract}
 \maketitle

\section{Introduction}

Historically, String theory suggests that our universe lives in a
higher dimensional space-time. Kaluza and Klein, in order to unify
electromagnetism with Einstein gravity, proposed that space-time
has more than three spatial dimensions \cite{1}. Later, in order
to explain some of the open questions in particle physics and
cosmology such as the hierarchy problem (which refers to the
difference in magnitude between the weak scale and the Planck
scale) and cosmological constant, higher dimensional space-time
with large extra dimensions were paid more attention. This line of
thinking led to the braneworld scenarios which are submanifolds
embedded in a higher dimensional space-time (bulk). In this
theory, it is supposed that our universe is like a membrane and
particles corresponding to electromagnetic, weak and strong
interactions are confined within it. Only gravitation and some
exotic matter (e.g., the dilaton field) could propagate in the
bulk. Serious research in the field of extra dimensions came with
the work of Arkani-Hamed, Dimopoulos, and Dvali who proposed the
large extra dimensions model, which lowers the energy scale of
quantum gravity to 1 TeV by localizing the standard model fields
to a 4-brane so that the hierarchy problem can be addressed
\cite{2, 3}. But, subsequently, it was shown that propagation of
gravity in the bulk is in contradiction with the observational
fact that four-dimensional gravity satisfies an inverse-square
Newtonian law. This problem was solved in a model proposed by
Randall and Sundrum (RS) by relaxing assumptions that our
four-dimensional universe is independent of the coordinates
defining the extra dimensions \cite{4, 5}. One can then show that
(even when the extra dimensions are infinitely large) gravity can
be localized near the 3-brane, and Newtonian gravity can be
restored at long distances. Indeed, RS considered the 3-brane (the
four-dimensional Minkowski space-time) embedded in the
five-dimensional anti-de Sitter space-time ($AdS_{5}$). They found
that there exists a massless graviton (0-mode) and massive
gravitons (Kaluza-Klein modes). The massless graviton reproduces
the Newtonian gravity on the 3-brane and Kaluza-Klein modes, which
are the effect of the existence of the higher-dimension, give a
correction to the Newtonian gravity \cite{6, 7}. Generally, this
model succeeds in the localization of gravity around the brane due
to the warping of the extra-dimension, but, these kinds of models
can be only treated as an approximation since any fundamental
theory would have a minimal length scale. Because in RS braneworld
scenarios, the brane is an infinitely thin object, the energy
density of the brane is a delta-like function with respect to the
fifth dimension coordinate. So this model is a very idealized
braneworld model. Recently, more realistic thick brane models were
investigated in higher dimensional space-time \cite{8}-\cite{12}.
In thick brane scenarios, the coupling between gravity and scalars
should be introduced. The presence of a scalar field makes the
warp function to behave smoothly.  In \cite{24}-\cite{34} some
properties of brane models were investigated: localization of
gravity, graviton ground state, stability.

This paper is organized as follows: In Sec. 2, the general
formalism of the braneworld scenario will be reviewed. Section 3
is devoted to introducing a new model and in Sec. 4 the stability
of the model is studied. In Sec. 5, the WD equation and its
solution will be investigated. Finally, in Sec. 6, concluding
remarks are presented.

\section{Thick Brane Formalism}

We consider a five-dimensional Einstein-scalar field theory with
the following action
\begin{equation}\label{ac}
S=\int d^{5}x \sqrt{g^{(5)}} \left( \dfrac{1}{4}R^{(5)} -\dfrac{1}{2}\partial_{A}\phi \partial^{A}\phi -V(\phi)\right).
\end{equation}
By variation with respect to the metric and scalar field, we get
\begin{equation}\label{eqmotion}
R_{A B}-\dfrac{1}{2}g_{A B}R^{(5)}=T_{AB}^{\phi} \hspace{0.5cm},\hspace{0.5cm} \square_{5}\phi=\dfrac{dV}{d\phi},
\end{equation}
where $ \square_{5}\phi=g^{AB}\phi _{;AB} $ and  $ T_{AB}^{\phi} $ is the 5-D energy-momentum tensor of the scalar field which is given by
\begin{equation}
T_{A B}=\partial_{A}\phi \partial_{B}\phi-g_{A B}\left( \dfrac{1}{2}\partial_{C}\phi \partial^{C}\phi+V(\phi)\right).
\end{equation}
The metric of a static five-dimensional space-time with the four-dimensional Poincare symmetry can be written as
\begin{equation}
ds_{5}^{2}=g_{A B} dx^{A} dx^{B}=a(w)^{2}\left(-dt^{2}+dx^{2}+dy^{2}+dz^{2} \right)+dw^{2},
\end{equation}
in which $ a(w) $ is the warp function. Assuming that the scalar
field $ \phi $ is a function of $ w $ only, the Einstein and
scalar equations are explicitly given by:
\begin{equation}\label{eq5}
(\mu,\nu):\hspace{1cm} 3\dfrac{a^{''}}{a}+3\dfrac{a^{'2}}{a^{2}}=-\dfrac{1}{2}\phi^{'2}-V,
\end{equation}
\begin{equation}\label{eq6}
(w,w):\hspace{2.5cm}6\dfrac{a^{'2}}{a^{2}}=\dfrac{1}{2}\phi^{'2}-V,
\end{equation}
\begin{equation}\label{eq711}
(scalar field):\hspace{1cm}\phi^{''}+4\dfrac{ a^{'}}{a}\phi^{'}=\dfrac{dV}{d\phi},
\end{equation}
where the prime denotes derivative with respect to $ w $. Only two
of the above equations are independent from each other. In order
to get first-order equations, sometimes, an auxiliary
superpotential is introduced which is related to $ V $ according
to \cite{8, 28, 29}
\begin{equation}\label{eq7}
V=-6 W(\phi)^{2}+\dfrac{9}{2}\left(\dfrac{dW(\phi)}{d\phi} \right)^{2}.
\end{equation}
Using the above equations (\ref{eq5}-\ref{eq711}) we get
\begin{equation}\label{eq8}
\phi^{'}=3 \dfrac{dW}{d\phi} \hspace{0.5cm},\hspace{0.5cm}\dfrac{a^{'}}{a}=-W(\phi).
\end{equation}
The energy density is then given by
\begin{equation}\label{eq9}
 T_{00}=\rho(w)=a(w)^{2}\left( \dfrac{1}{2}(\dfrac{d \phi}{d w})^{2}+V(\phi)\right) =\dfrac{d}{dw}\left(3 W(\phi) a(w)^{2}\right).
\end{equation}
For a given metric function, one can determine the scalar field
profile, potential and superpotential, using the above equations.

\section{The LambertW ($\mathcal{W}$) Model}

\begin{center}
\begin{figure}[H] \hspace{4cm}\includegraphics[width=8.cm]{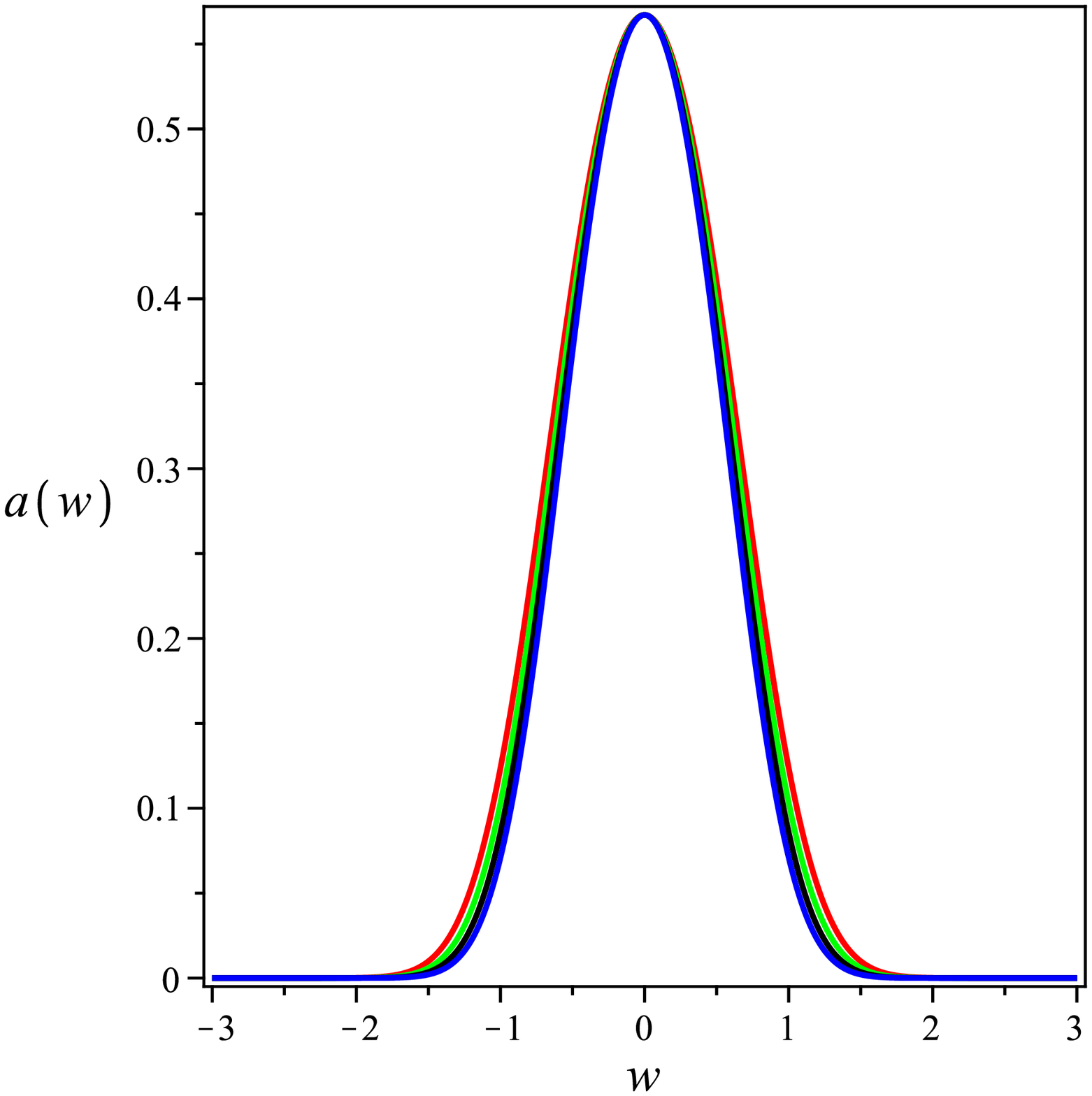}\vspace{0.1cm}\caption{\label{fig0} \small
 The behavior of warp factor in terms of $ w $ for $\alpha =\textcolor{red}{1}, \textcolor{green}{1.1}, 1.2,\textcolor{blue}{1.3}$ (downwards).}
\end{figure}
\end{center}

\begin{center}
\begin{figure}[H] \hspace{4cm}\includegraphics[width=8.cm]{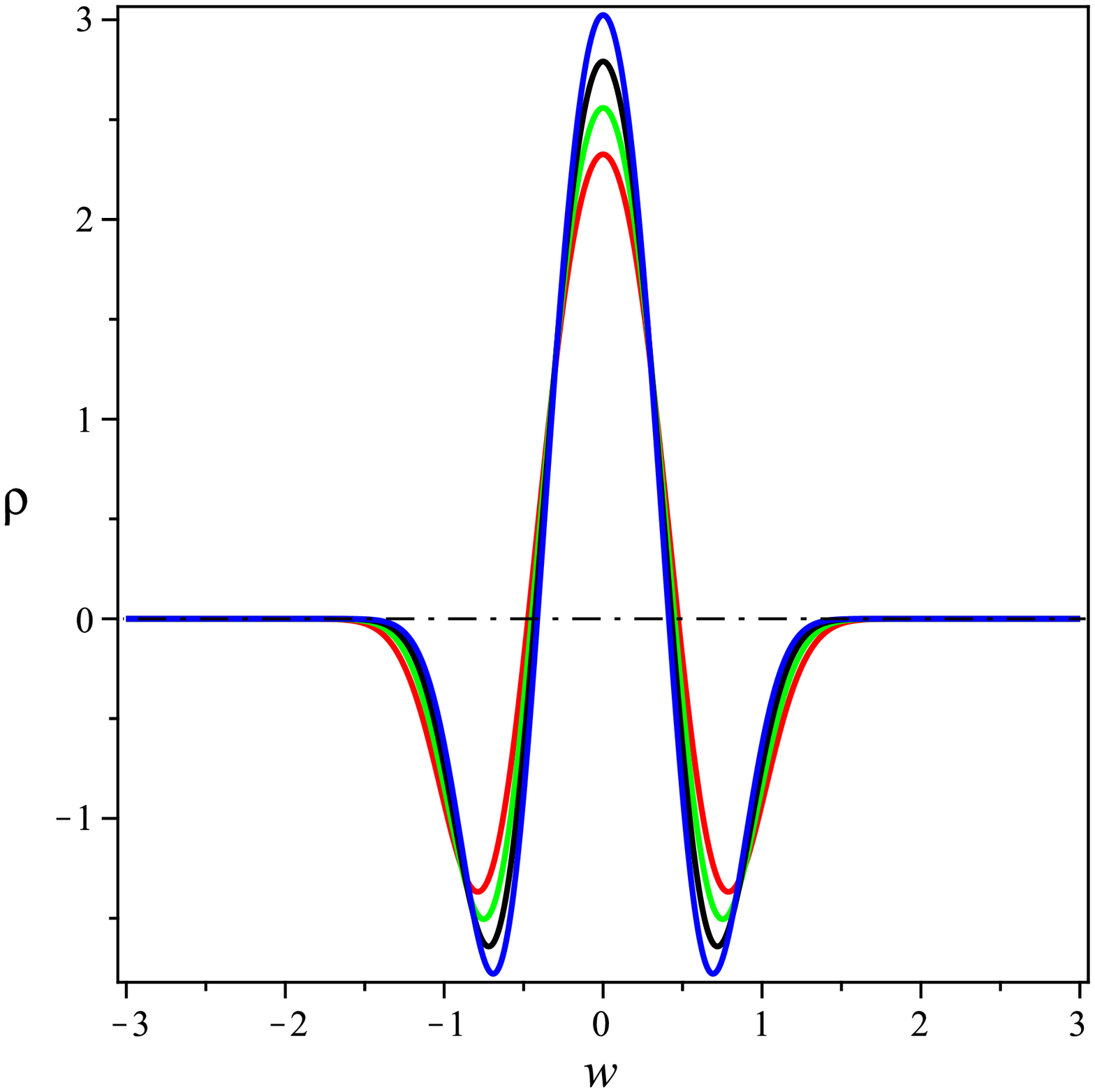}\vspace{0.1cm}\caption{\label{fig2} \small
 The behavior of the energy density ($ T_{00}=\rho $) in terms of $ \omega $ at $\alpha =\textcolor{red}{1}, \textcolor{green}{1.1}, 1.2, \textcolor{blue}{1.3}$.}
\end{figure}
\end{center}

We study the case of a (conformally) flat brane, with the line
element
\begin{equation}\label{om}
ds^{2}=a(w)^{2}\eta_{\mu \nu} dx^{\mu} dx^{\nu}+dw^{2} \hspace{0.5cm},\hspace{0.5cm} a(w)=\mathcal{W}\left(\dfrac{1}{1+\alpha w^{2}}\right)^{1+\alpha w^{2}},
\end{equation}
where $ \eta_{\mu \nu} $ is the 4-dimensional Minkowski metric
with signature (-,+,+,+) and $ \alpha $ is a positive constant
parameter with dimension $1/[length]^2$ that controls the
localization of the brane. Moreover, as one can see later in the
definition of Einstein tensor, this quantity plays the role of
cosmological constant on the brane. The warp factor has the $
Z_{2} $ symmetry $ a(w)=a(-w) $, and is plotted in Fig.
\ref{fig0}. The asymptotic behavior of the metric is
\begin{equation}\label{eqapp12}
ds^{2}\approx \dfrac{0.02}{(\alpha^2 w^{4})^{\alpha w^{2}+1}}\eta_{\mu \nu}dx^{\mu}dx^{\nu}+dw^{2} \hspace{0.5cm},\hspace{0.5cm}w \rightarrow \infty,
\end{equation}
and near the brane
\begin{equation}
ds^{2}\approx (0.3-0.8\alpha w^2)\eta_{\mu \nu}dx^{\mu}dx^{\nu}+dw^{2} \hspace{0.5cm},\hspace{0.5cm}w \rightarrow 0.
\end{equation}
 Also, one can calculate the energy density from Eq. (\ref{eq9}), which is plotted in Fig.
 \ref{fig2}. The asymptotic and near brane behavior of the energy density are
 \begin{align}
 T_{00} \approx & -\dfrac{0.44}{w^2 (\alpha^2 w^4)^{(\alpha w^2)}}(1+\ln(\alpha w^2))^2 \hspace{0.5cm},\hspace{0.5cm}w \rightarrow \infty\nonumber\\
 T_{00} \approx & 2.33 \alpha -12.3\alpha^2 w^2+3.34\alpha^3 w^4 \hspace{0.5cm},\hspace{0.5cm}w \rightarrow 0.
  \end{align}


\begin{center}
\begin{figure}[H] \hspace{4cm}\includegraphics[width=8.cm]{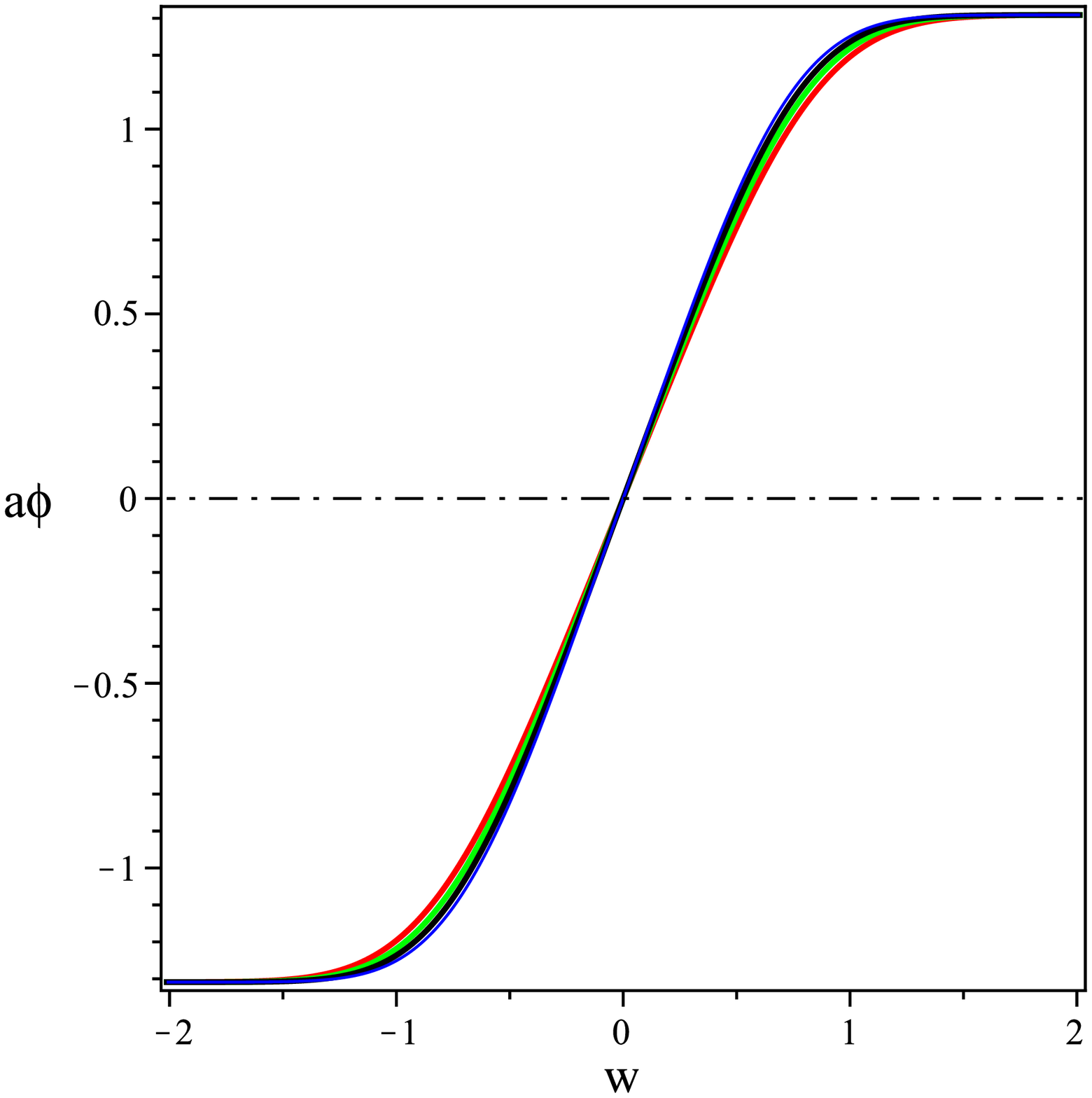}\vspace{0.1cm}\caption{\label{figphi} \small
 $ a \phi$ as a function of $w$ at $\alpha =\textcolor{red}{1}, \textcolor{green}{1.1}, 1.2, \textcolor{blue}{1.3}$.}
\end{figure}
\end{center}

Considering Figs. \ref{fig0} and \ref{fig2}, one finds that $
\alpha $ is a parameter that controls both the localization width
of the brane and the amplitude of the localization of energy
density on the brane. As this parameter increase, the warp factor
will be more localized and the amplitude of energy density
decreases.

\begin{center}
\begin{figure}[H] \hspace{4cm}\includegraphics[width=8.cm]{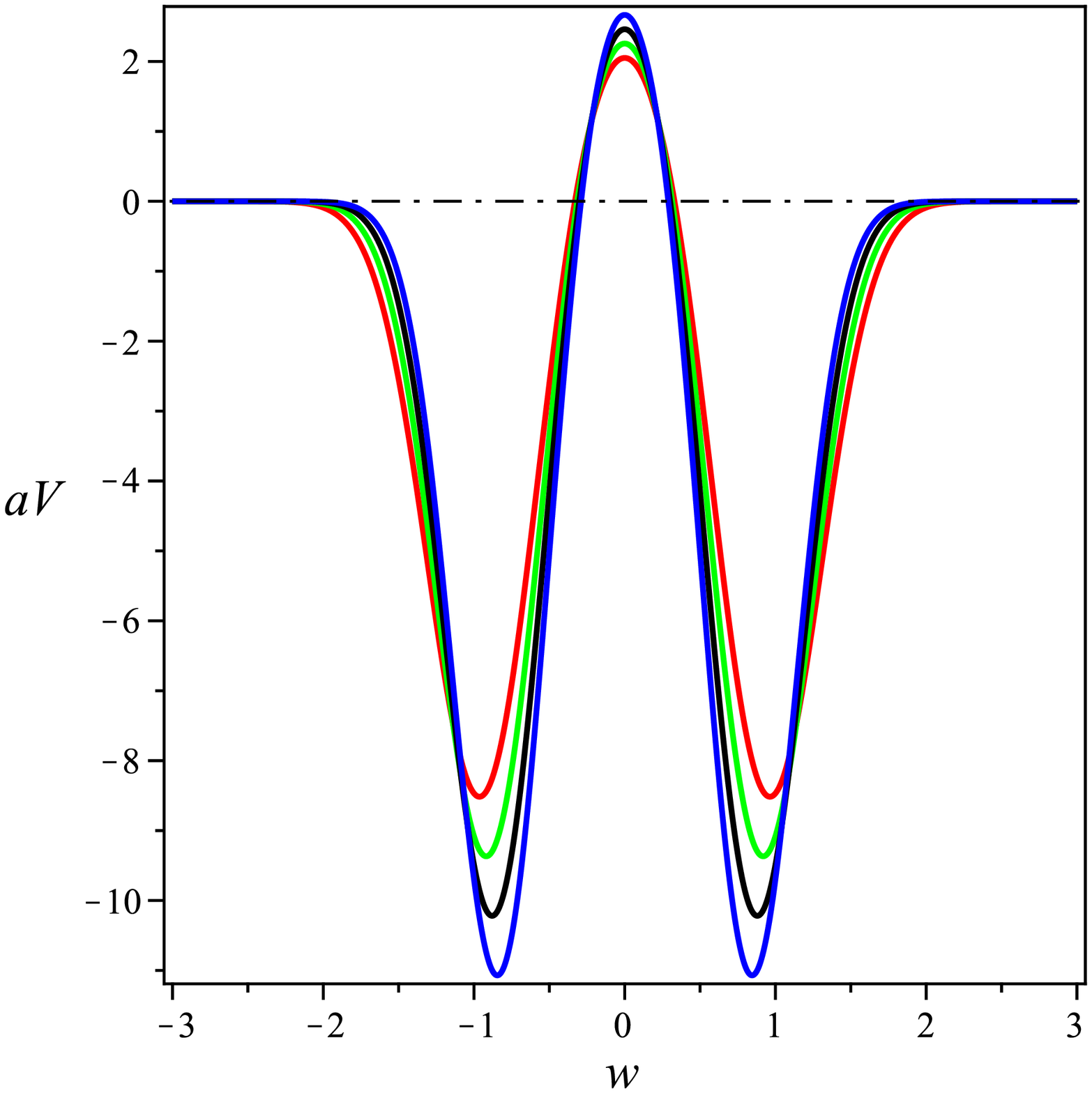}\vspace{0.1cm}\caption{\label{fig:ri} \small
$aV$ as a function of the $ w $ at  $ \alpha =\textcolor{red}{1}, \textcolor{green}{1.1}, 1.2, \textcolor{blue}{1.3}$.}
\end{figure}
\end{center}

From Eq. (\ref{eq8}), it will be easy to calculate:
\begin{equation}
W=-2\alpha w\left(A-\dfrac{1}{1+A}\right),
\end{equation}
\begin{equation}\label{eq12}
\left(\dfrac{d\phi}{dw}\right)^{2}=-6\alpha\ln A+\dfrac{12\alpha^2 w^2}{(1+\alpha w^2)\left(1+A\right)}
+\dfrac{6\alpha}{1+A}+\dfrac{12\alpha^2 w^{2}A}{(1+\alpha w^2)\left(1+A\right)^3},
\end{equation}
where $A=\mathcal{W}\left(\dfrac{1}{1+\alpha w^{2}}\right)$. By
using of ODE plot method, one can obtain the behavior of the
scalar field from Eq. (\ref{eq12}). As can be seen from Fig.
\ref{figphi}, the scalar field has mirror symmetry ($
\phi(w)=-\phi(-w) $). The potential of this system from Eq.
(\ref{eq7}) as a function of fifth dimension is given by:
\begin{multline}\label{eqpotential}
V(w)=-24\alpha^2 w^2\left(\ln A-\dfrac{1}{1+A}\right)^2-
3\alpha\left(\ln A-\dfrac{1}{1+A}\right)+\dfrac{6\alpha^2
w^2}{(1+A)(1+\alpha w^2)}\left(1+\dfrac{A}{(1+A)^2}\right).
\end{multline}

This potential is plotted in the Fig. \ref{fig:ri}. It can be seen
that the potential is an even function of $ w $ ($ V(w)=V(-w) $).
In order to obtain $V(\phi)$, we need to have $\phi(w)$.
Considering Eq. (\ref{eq12}), it is obvious that one cannot obtain
an analytic relation for $\phi(w)$. However we have obtained
approximation solutions for $\phi$ for both small and large $w$,
as follows
\begin{align}
\phi \approx & \; 7.23\alpha w+\mathcal{O}(w^3),\hspace{1.5cm}w \rightarrow 0\nonumber\\
\phi \approx & \; 6\alpha w(1+\ln(\alpha w^2))+\mathcal{O}\left(\dfrac{1}{w}\right),\hspace{0.5cm}w \rightarrow \infty
\end{align}
Now, by obtaining the function $w(\phi)$, one can get the scalar
potential as follow
\begin{align}
V(\phi)\approx \; & 3.61\alpha -0.53\phi^2 +\mathcal{O}(\phi^4),\hspace{0.5cm}w \rightarrow 0 \nonumber\\
V(\phi)\approx & \; 6 \alpha-0.667 \phi^2
+\mathcal{O}\left(\dfrac{1}{\phi^2}\right),\hspace{0.5cm}w
\rightarrow \infty
\end{align}
in which confirm that the potential is an even function of scalar
field ($V(\phi)=V(-\phi)$).

Here, we are going to investigate the curvature behavior of the
spacetime. The Ricci and Kretschmann scalars are given as
\begin{align}
R&=\nonumber\\
&-\dfrac{16\alpha \left[ (1+\alpha w^2)(1+A)^2(1+A+10\alpha A w^2)\ln(A)+5\alpha^2 w^4(1+A)-\alpha w^2(3A^2+3A-2)-(1+A)^2\right]}{(1+A)^3(1+\alpha w^2)},
\end{align}
and
\begin{align}
K&=R_{a b c d }R^{a b c d}=\nonumber\\
&+\dfrac{128\alpha^2 \ln(A)(20\alpha^3 w^6 A^4+26\alpha^2 w^4 A^4+13\alpha w^2 A^3+7\alpha w^2 A^4+24\alpha^2 w^4 A^3-18\alpha^2 w^4 A^2)}{(1+A)^4(1+\alpha w^2)}\nonumber\\
&+\dfrac{128\alpha^2 \ln(A)(-8 A\alpha^2 w^4-7 A\alpha w^2+20\alpha^3 A^3 w^6+A^4+3A^3+3A^2+\alpha A^2 w^2+A-2\alpha w^2)}{(1+A)^4(1+\alpha w^2)}\nonumber\\
&+\dfrac{64\alpha^2(1+3\alpha^2 w^4+9\alpha^2 w^4 A^4+24\alpha w^2 A^3+6\alpha w^2 A^4+32\alpha^2 w^4 A^3+36\alpha^2 w^4 A^2+12A\alpha^2 w^4)}{(1+A)^6(1+\alpha w^2)^2}\nonumber\\
&+\dfrac{64\alpha^2(+16A\alpha w^2-24\alpha^3 A^2 w^6+10\alpha^4 A^2 w^8+20\alpha^4 w^8 A-12\alpha^3 w^6 A^3+A^4+4A^3+6A^2)}{(1+A)^6(1+\alpha w^2)^2}\nonumber\\
&+\dfrac{64\alpha^2(-4\alpha^3 w^6 A+32\alpha w^2 A^2+10\alpha^4 w^8+8\alpha^3 w^6+4A+2\alpha w^2)}{(1+A)^6(1+\alpha w^2)^2}.
\end{align}
In the limit of $ w\rightarrow \pm\infty $ and $ w \rightarrow 0
$, we find
\begin{equation}
\lim_{w\to \pm \infty} R=-\infty,
\end{equation}
\begin{equation}
\lim_{w\to  0} R=19.3\alpha,
\end{equation}
 \begin{equation}
 \lim_{w\to  0} K=92.97 \alpha^{2},
 \end{equation}
 \begin{equation}
 \lim_{w\to  \pm \infty} K=\infty.
 \end{equation}

Notably, divergence values of curvature scalars for $\omega
\rightarrow \infty$ do not have a physical interpretation. In
order to overcome such a problem, we applied a constraint on the
free (positive) parameter $\alpha$ in such a way that for $\omega
\rightarrow \infty$ we have finite value for $\alpha \omega^2$ (we
set this finite value to one without loss of generality).
Considering the mentioned limitation, one finds an asymptotically
flat $5-$dimensional spacetime (as $\omega \rightarrow \infty$).

Besides, the components of the Einstein tensor can be written as
\begin{align}
G^{\mu}_{\nu}&=\nonumber\\
&\dfrac{6\alpha \left[(1+\alpha w^2)(1+A)^2(1+A+8\alpha A w^2)
\ln(A)+4\alpha^2 w^4 (1+A)-\alpha
w^2(3A^2+4A-1)-(1+A)^2\right]\delta^{\mu}_{\nu}}{(1+A)^3(1+\alpha
w^2)},
\end{align}
and
\begin{equation}
G^{w}_{w}=\dfrac{24\alpha^2 w^2((1+A)\ln(A)-1)^2}{(1+A)^2},
\end{equation}
where in the limits $w \rightarrow 0$, one can find
\begin{equation}
\lim_{w\to  0} G^{\mu}_{\nu}=-7.23 \alpha \delta^{\mu}_{\nu},
\label{Cos}
\end{equation}
\begin{equation}
\lim_{w\to  0} G^{w}_{w}=0.
\end{equation}
So, one can interpret the right hand side of Eq. (\ref{Cos}) as
the cosmological constant on the brane (immersed in
$5-$dimensional spacetime), i.e., $\Lambda=7.23 \alpha$ and since
$\alpha$ is positive, the cosmological constant of the brane would
be positive for this model. Indeed, regarding Eq. (\ref{om}), it
is clear that the four dimensional brane is conformally flat.
However, the $5-$dimensional spacetime is not flat at all. It is
notable that for the limiting $\omega \rightarrow 0$, we
investigated the geometrical properties of the brane immersed in
$5-$dimensional spacetime. In other words, although we use the
limit $\omega \rightarrow 0$ to localize on the brane, it is
notable that such a brane in submanifold of $5-$dimensional
spacetime. So after calculating the Einstein and Ricci tensors of
$5-$dimensional spacetime, we find the trace of cosmological
constant near the brane and $5-$dimensional spacetime behaves like
a dS space for $\omega \rightarrow 0$.

\section{Stability}
\begin{center}
\begin{figure}[H] \hspace{4cm}\includegraphics[width=8.cm]{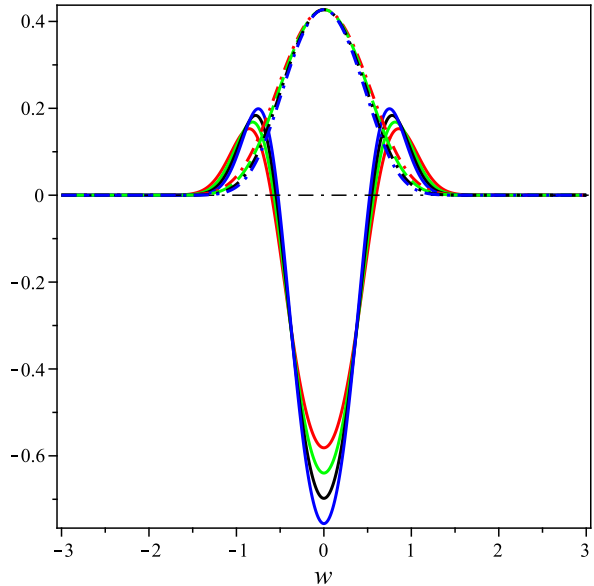}\vspace{0.1cm}\caption{\label{fig:pot} \small
 The behavior of zero-mode wave function (dotted-dashed lines) and effective potential (solid lines) in terms of $ w $ for $\alpha =\textcolor{red}{1}, \textcolor{green}{1.1}, 1.2,\textcolor{blue}{1.3}$ (downwards).}
\end{figure}
\end{center}

Another general feature concerns the stability of the gravity
sector of the braneworld model. Here, we shall consider linear
perturbations of the metric in the following form \cite{30}
\begin{equation}
ds^{2}=a(w)^{2}(\eta_{\mu \nu}+\varepsilon h_{\mu \nu})dx^{\mu}dx^{\nu}+dw^{2}
\end{equation}
where $ \varepsilon h_{\mu \nu} $ is a small perturbation around
the Minkowski metric. By considering the transverse and traceless
gauge and re-defining the $ w $-coordinate as $ dw=a(w) dz $, the
corresponding Schr${\rm \ddot{o}} $dinger-like equation takes the
following form
\begin{equation}
-\dfrac{d^{2}\Psi(z)}{dz^{2}}+U(z)\Psi(z)=\lambda^{2}\Psi(z),
\end{equation}
where $ \lambda^{2} $ accounts for the 4D mass of the excited
Kaluza-Klein gravitational modes and $\Psi(z)  $ is a wave
function. The stability potential is given by
\begin{equation}
U(z)=\dfrac{3}{2}\dfrac{\ddot{a}(z)}{a(z)}+\dfrac{3}{4}\dfrac{\dot{a}(z)^{2}}{a(z)^{2}}=\dfrac{3}{4}\left(a^{''}(w)a(w)+\dfrac{1}{4}a^{'2}(w)\right).
\end{equation}
 The asymptotic behavior of the effective potential for our model is
\begin{equation}
U(w\gg 0)=\dfrac{0.27\ln(\alpha w)}{(\alpha w^{2})^{2\alpha w^2}w^2}+\mathcal{O}( \dfrac{1}{w^{4}}),
\end{equation}
and near the center of potential
\begin{equation}
U(w\approx0)=-0.59\alpha+2.017\alpha^2 w^2+\mathcal{O}(w^3).
\end{equation}
\\
This potential is plotted in Fig. \ref{fig:pot} for different
values of $ \alpha $. As one can see, this potential has only one
minimum at $ w = 0$ which implies stability. Also, the effective
potential shows that gravitons are localized around the brane. We
know that the solution of this Schr${\rm \ddot{o}} $dinger
equation at $ w=0 $, represents the coupling of the massive modes
with the matter on the brane. The zero mode is $ \Psi_{0}\propto
a(z)^{\frac{3}{2}} $ \cite{30,33}

\begin{center}
\begin{figure}[H] \hspace{4cm}\includegraphics[width=8.cm]{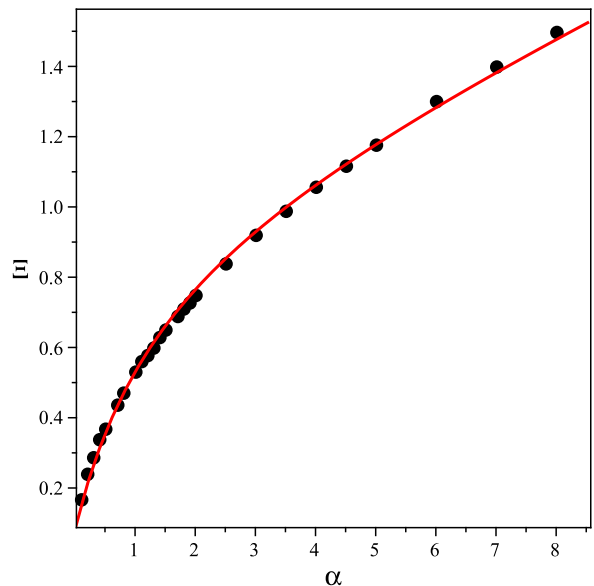}\vspace{0.1cm}\caption{\label{figX} \small
 The behavior of $\Xi$ in terms of $ \alpha $.}
\end{figure}
\end{center}

In order to ensure stability, it is important to check the
nomalizability of $ \Psi_{0}^{2} $. Normalizability is connected
with the asymptotic behavior of the potential of the Schr${\rm
\ddot{o}} $dinger equation. If $ U(z)>0 $ as $ |z| \rightarrow
\infty $, then $\Psi_{0}(z)  $ is always normalizable \cite{22}.
In the following, we want to obtain the correction from the
massive modes of Kaluza-Klein for the four-dimensional
gravitational coupling \cite{10}. First, we obtain Newtonian
coupling by using of zero-modes as follows:
\begin{equation}
G_{4}\sim \dfrac{1}{M_{\ast}^{3}}\dfrac{\Psi_{0}^{2}(0)}{<\Psi_{0}\mid \Psi_{0}>}\equiv \dfrac{1}{M_{\ast}^{3}} \Xi,
\end{equation}
where $ M_{\ast} $ is the five-dimensional fundamental scale and
$\Xi \equiv \dfrac{\Psi_{0}^{2}(0)}{<\Psi_{0}\mid \Psi_{0}>}$. In
order to obtain a correction for $G_{4}$ from massive Kaluza-Klein
modes, we have used numerical method. In Fig. (\ref{figX}), we
have plotted the behavior of $\Xi$ in terms of $\alpha$. Then, we
have fitted the best function on the numeric data. So, the best
function is as follows:
\begin{equation}
\Xi \sim \dfrac{0.08(\alpha+14.07)(\alpha+0.11)}{\alpha+1.44}\sim 1+0.08 \alpha+...\hspace{1cm} \alpha\gg 1
\end{equation}

The effective potential between two point-like sources of mass $
M_{1} $  and $ M_{2} $ is from the contribution of the zero mode
and the continuum KK modes that can be expressed as \cite{10}
\begin{equation}\label{eqpot}
U(r)=\dfrac{M_1 M_2}{r} (G_4 +\delta G_{4})=\dfrac{M_{1}M_{2}}{r}M_*^{-3}(\Xi+\int_{\lambda_{0}}^\infty d \lambda e^{-\lambda r}|\Psi_{\lambda}(0)|^{2}),
\end{equation}\\
in which the continuous spectrum starts at $ \lambda_{0} $. In
order to get the corrected Newtonian potential, we have obtained
the wave function near $w=0$ as follows:
\begin{equation}
\Psi_{\lambda}(w)\approx C_{1}\sin(3.1\sqrt{(0.2\alpha+0.3\lambda^2)}w)+C_{2}\cos(3.1\sqrt{(0.2\alpha+0.3\lambda^2)}w)
\end{equation}
then $\Psi_{\lambda}(0)=C_{2}$ and by inserting it into Eq. (\ref{eqpot}), one gets
\begin{equation}
U(r)=\dfrac{M_{1}M_{2}}{r}M_{\ast}^{-3}(\Xi+C_{2}^{2} { \dfrac{e^{-{\lambda_{0}} r}}{r})},
\end{equation}
in order to have a correction to the Newtonian potential
$C_{2}\neq 0$. From the above discussion, it can be seen that for
large distance $ r $ between two point-like sources, the
correction of gravitational potential $ \dfrac{e^{-{\lambda_{0}}
r}}{r^{2}} $ is very small compared with the Newtonian potential $
\frac{1}{r} $, because the contribution of massive Kaluza-Klein
modes is at large distances. However, when $ r $ is the order of $
\lambda_{0}^{-1} $ or smaller, the correction
$ \dfrac{e^{-{\lambda_{0}} r}}{r^{2}} \sim \dfrac{1}{r^{2}}$ becomes important in the effective Newtonian potential \cite{34, 35}.\\

\section{Wheeler-DeWitt equation}

The first approach to describe the universe based on the
application of the quantum theory was presented in 1960 by Wheeler
\cite{13} and DeWitt \cite{14}. They proposed a quantum gravity
equation to describe the wave function of the universe, which is
known as the Wheeler-DeWitt (WD) equation. This equation is
analogous to a zero-energy Schr${\rm \ddot{o}} $dinger equation in
which the Hamiltonian could contain the gravitational field and
scalar fields. But, one of the most important problems for solving
the WD equation is the subject of initial conditions. Unlike a
classical system, for cosmological models, there are no external
initial conditions  because there is no external time parameter to
the universe. In order to solve this problem, two different
approaches are used: the Hartle-Hawking no boundary
\cite{15}-\cite{18} and the Vilenkin tunneling proposal
\cite{19}-\cite{22}. The first proposal  is that the wave function
of the Universe is given by a path integral over compact Euclidean
geometries so that this universe has no boundary in this space.
The second one states that the universe spontaneously nucleates
and then evolves along the lines of an inflationary scenario. The
mathematical description of this approach is closely analogous to
that of quantum tunneling through a potential barrier. In fact,
only the outgoing modes of the wave function should be taken at
the singular boundary of superspace \cite{22, 23}.

In order to investigate the canonical quantization of the brane,
one should promote the metric $ g_{ij} $, the conjugate momenta $
\Pi_{ij} $, the Hamiltonian density $ H $ and the momentum density
$ H_{i} $ to quantum operators satisfying canonical commutation
relations. We begin with the metric
\begin{equation}\label{eqmetric}
ds^{2}=a(w)^{2}\left(-dt^{2}+\dfrac{b(t)^{2}}{1-\kappa r^2}dr^{2}+b(t)^{2}r^{2}d \theta^{2}+b(t)^{2}r^{2} sin(\theta)^{2} d \phi^{2} \right)+dw^{2},
\end{equation}
where $ b(t) $ is the brane scale factor. Ricci scalar for this metric is given by
\begin{equation}
R^{(5)}=-\dfrac{2(4a(w)b(t)^{2}a^{''}(w)+6b(t)^{2}a^{'}(w)^{2}-3b(t)\ddot{b}(t)-3\dot{b}(t)^{2}-3\kappa)}{a(w)^{2}b(t)^{2}},
\end{equation}
where dot and prim are derivatives with respect to $t$ and $w$, respectively.
By using Eq. (\ref{ac}), the Lagrangian is
\begin{multline}\label{ll}
L=2a(w)^{2}b(t)(-3\dot{b}^{2}-6b(t)^{2}a^{'}(w)^{2}+4a(w)b(t)^{2}a^{''}(w))+\dfrac{1}{2}a(w)^{2}b(t)^{3}\dot{\varphi}(w,t)^{2}-\\
\dfrac{1}{2}a(w)^{4}b(t)^{3}\varphi^{'}(w,t)^{2}-a(w)^4b(t)^{3}V(\varphi)+6a(w)^{2}b(t)\kappa.\hspace{6cm}
\end{multline} \\
The momenta conjugate to $ b $  and $ \varphi $ are
 \begin{equation}
 \Pi_{b}=\dfrac{\partial L}{\partial \dot{b}}=-12a(w)^{2}b(t)\dot{b} \hspace{0.5cm},\hspace{0.5cm}\Pi_{\varphi}=\dfrac{\partial L}{\partial \dot{\varphi}}=a(w)^{2}b(t)^{3}\dot{\varphi}.
 \end{equation}
 The Hamiltonian constraint is therefore
\begin{equation}
H=\Pi_{b}\dot{b}+\Pi_{\varphi}\dot{\varphi}-L=-\Pi_{b}^{2}+\dfrac{12}{b(t)^{2}}\Pi_{\varphi}^{2}+U(a,b,\varphi)=0,
\end{equation}
 where $ U(a,b,\varphi) $ is
\begin{equation}\label{eqsuperpot}
U(a,b,\varphi)=96a^{4}b^{4}(3a^{'2}+2aa^{''})+24a^{6}b^{4}(\dfrac{1}{2}\varphi^{'2}+V(\varphi))-144a^{4}b^{2}\kappa.
\end{equation}

Making the replacement $  \Pi_{b}  \rightarrow
-i\dfrac{\partial}{\partial b}$ and  $  \Pi_{\varphi}  \rightarrow
-i\dfrac{\partial}{\partial \varphi}$ and imposing $ H\Psi=0 $
results in the following WD equation
 \begin{equation}
 \left( \dfrac{\partial^{2}}{\partial b^{2}}-\dfrac{12}{b^{2}}\dfrac{\partial^{2}}{\partial \varphi^{2}}+U(a,b,\varphi)\right)\Psi=0.
 \end{equation}
In order to solve the WD equation we follow the separation of variable method. Using
 \begin{equation}
 \Psi(b, \varphi)= \Phi(\varphi) B(b),
 \end{equation}
 the WD equation becomes
 \begin{equation}\label{eq41}
 \dfrac{1}{B}\dfrac{d^{2}B}{db^{2}}-\dfrac{12}{\Phi b^{2}}\dfrac{d^{2}\Phi}{d\varphi^{2}}+\nu b^{4}-\varrho b^{2} =0,
  \end{equation}
where by using of Eq. (\ref{eq12}), (\ref{eqpotential}) and
(\ref{eqsuperpot}),
$\nu=(96a^{4}(3a^{'2}+2aa^{''})+24a^{6}(\frac{1}{2}\varphi^{'2}+V(\varphi)))\mid_{w=0}=-15.1\alpha
$ and $ \varrho=144a(w=0)^{4}\kappa=14.898 \kappa $. Here, we
assumed $ \varphi(w,t)=\varphi(t) $ and $V(\varphi)=0$. We thus
obtain the following equations
 \begin{equation} \label{eq45}
 \dfrac{d^{2}\Phi}{d\varphi^{2}}-m \Phi=0,
 \end{equation}
 \begin{equation}\label{eq43}
 \dfrac{d^{2}B}{db^{2}}+(\nu b^{4}-\varrho b^{2}-\dfrac{12m}{b^{2}})B=0.
 \end{equation}
where $ m $ is a separation constant. By solving equation (\ref{eq45}) one obtains
 \begin{equation}
 \Phi(\varphi)=c_{1}e^{\sqrt{m}\varphi}+c_{2}e^{-\sqrt{m}\varphi}.
 \end{equation}
If in Eq. (\ref{eq43}) $ b\rightarrow 0 $, one obtains
 \begin{equation}
 \dfrac{d^{2}B}{db^{2}}-\dfrac{12m}{b^{2}}B=0,
 \end{equation}
which has the following solution
 \begin{equation}
B(b \rightarrow 0) \approx c_{1} \sqrt{{b}}^{(1+\sqrt {48m+1})}+c_{2} \sqrt{{b}}^{(1-\sqrt {48m+1})},
\end{equation}
in the limit $ b\rightarrow \infty $ Eq. (\ref{eq43}) becomes
\begin{equation}
 \dfrac{d^{2}B}{db^{2}}+(\nu b^{4}-\varrho b^{2})B=0,
 \end{equation}
which has the following solution:
 \begin{equation}
B(b \rightarrow \infty ) \approx c_{1} e^{f(b,\kappa)}HenuT(\lambda,\delta,\eta, \zeta b)+c_{2}
e^{-f(b,\kappa)}HenuT(\lambda,\delta,\eta, -\zeta b),
 \end{equation}
 where
\begin{eqnarray}
  f(b,\kappa)&=&\dfrac{(-0.33+0.11 \times 10^{-9}I)(-1.5\varrho+\nu b^2)b}{\sqrt{-\nu}},\nonumber \\
  \lambda&=&\dfrac{-(0.16+0.28I)\varrho^2}{(-\nu)^{(\frac{4}{3})}},\nonumber \\
  \delta&=&0,\nonumber \\
  \eta&=&\dfrac{(-0.57+0.99I)\varrho}{(-\nu)^{\frac{2}{3}}}, \nonumber \\
  \zeta&=&(0.44+0.76I)(-\nu)^{\frac{1}{6}}b.\nonumber
\end{eqnarray}

According to Eq. (\ref{eq41}), the effective potential is
\begin{equation}\label{eq52}
 U(b,w=0)=\nu b^{4}-\varrho b^{2}.
\end{equation}

In order to investigate the WD equation, we need to know the
properties of the superpotential $ U(b,w) $. The superpotential
may have a maximum, necessary for quantum tunneling. For $ w=0 $
and $ b\gg 0 $ the superpotential consists of two terms, a curvature
term $ \varrho b^{2} $ and the term $ \nu b^{4} $. Since $\nu <0 $
$ (\alpha>0) $ and $ \varrho <0 $ or $(\kappa <0) $, we have
quantum tunneling.

In the following, in order to more study the case of $\kappa=-1$,
we study cosmology in brane. So, By using of our ansatz metric
(\ref{eqmetric}), the explicit form of Einstein equation and the
equation of motion for $ \phi $ resulting from the action
(\ref{eqmotion}) are \cite{Ahmed:2013lea}
\begin{equation}\label{eq40}
(t,t):\hspace{2cm} \dfrac{3}{a^{2}}\left( a^{' 2}+a a^{''}-\dfrac{\dot{b}^2}{ b^2}-\dfrac{\kappa}{ b^2}\right)  =\dfrac{1}{2}\dot{\phi}^2-\dfrac{1}{2}\phi^{'2}-V,
\end{equation}
\begin{equation}\label{eq50}
(i,j):\hspace{0.5cm} \dfrac{1}{a^2}\left(-2\dfrac{\ddot{b}}{b}-\dfrac{\dot{b}^2}{b^2}-\dfrac{\kappa}{b^2}+3a^{' 2}+3aa^{''}\right)=-\dfrac{1}{2}\dot{\phi}^2-\dfrac{1}{2}\phi^{'2}-V,
\end{equation}
\begin{equation}\label{eq60}
(w,w):\hspace{2cm}\dfrac{3}{a^{2}}\left( -\dfrac{\ddot{b}}{b}+2a^{' 2}-\dfrac{\dot{b}^2}{b^2}-\dfrac{\kappa}{b^{2}}\right) =-\dfrac{1}{2}\dot{\phi}^2+\dfrac{1}{2}\phi^{'2}-V,
\end{equation}
\begin{equation}\label{eq71}
(scalar field):\hspace{3cm}\ddot{\phi}+3\dfrac{\dot{b}}{b}\dot{\phi}+a^2\dfrac{\partial V}{\partial \phi}-a^{2}\phi^{''}+4a^{'}a\phi^{'}=0,
\end{equation}
in the case of $\phi(w,t)=\phi(w)$, the equations simplify as follow:

\begin{equation}\label{eq44}
(t,t):\hspace{1cm} \dfrac{\dot{b}^2}{ b^2}+\dfrac{\kappa}{ b^2}=\dfrac{a^{2}}{3}\left[3\left(\dfrac{a^{' 2}}{a^{2}}+\dfrac{a^{''}}{a}\right)+\dfrac{1}{2}\phi^{'2}+V\right],
\end{equation}
\begin{equation}\label{eq55}
(i,j):\hspace{0.5cm} 2\dfrac{\ddot{b}}{b}+\dfrac{\dot{b}^2}{b^2}+\dfrac{\kappa}{b^2}=a^2\left[3\left(\dfrac{a^{' 2}}{a^{2}}+\dfrac{a^{''}}{a}\right)+\dfrac{1}{2}\phi^{'2}+V\right],
\end{equation}
\begin{equation}\label{eq66}
(w,w):\hspace{2cm} \dfrac{\ddot{b}}{b}+\dfrac{\dot{b}^2}{b^2}+\dfrac{\kappa}{b^{2}} =\dfrac{a^2}{3}\left[6\dfrac{a^{' 2}}{a^2}-\dfrac{1}{2}\phi^{'2}+V\right],
\end{equation}
\begin{equation}\label{eq77}
(scalar field):\hspace{4cm} \phi^{''}-4\dfrac{a^{'}}{a} \phi^{'}-\dfrac{\partial V}{\partial \phi}=0,
\end{equation}
where the left(right) hand sides depend only on $t$ ($w$). We then obtain the following
set of equations for $b(t)$:
\begin{equation}\label{eqfre1}
\dfrac{\dot{b}^2}{b^2}+\dfrac{\kappa}{b^2}=C_{t},
\end{equation}
\begin{equation}\label{eqfre2}
\dfrac{2\ddot{b}}{b}+\dfrac{\dot{b}^2}{b^2}+\dfrac{\kappa}{b^2}=C_{x},
\end{equation}
\begin{equation}\label{eqfre3}
\dfrac{\ddot{b}}{b}+\dfrac{\dot{b}^2}{b^2}+\dfrac{\kappa}{b^2}=C_{w},
\end{equation}
where $C_{t,x,w}$ are constants. It is easy to see that in order for the first two equations
to be consistent with the third one it is necessary that
\begin{equation}
C_{w}=\dfrac{C_{x}+C_{t}}{2}.
\end{equation}
On the other hand, from the right hand sides of equations one obtains for the
$w$-dependent functions the following equations
\begin{equation}
C_{x}=3C_{t}
\end{equation}
so that all the constants can be expressed in terms $ C_{w} $
\begin{equation}
C_{t}=\dfrac{1}{2}C_{w}=\dfrac{1}{2}\lambda,\hspace{1cm}C_{x}=\dfrac{3}{2}C_{w}=\dfrac{3}{2}\lambda
\end{equation}
where $\lambda$ is a constant. Then by combination of Eqs.
(\ref{eqfre1})-(\ref{eqfre3}), one finds the following
differential equations
\begin{equation}\label{eqfree}
\dfrac{\ddot{b}}{b}-\dfrac{\dot{b}^2}{b^2}-\dfrac{\kappa}{b^2}=0,
\end{equation}
\begin{equation}\label{eqfre}
\dfrac{\dot{b}^2}{b^2}+\dfrac{\kappa}{b^2}-\dfrac{\lambda}{2}=0.
\end{equation}
By solving Eq. (\ref{eqfree}), one can obtain
\begin{equation}\label{eqscal}
b(t)=\dfrac{c_{1}}{2}\left(e^{\dfrac{t+c_{2}}{c_{1}}}+\kappa e^{-\dfrac{t+c_{2}}{c_{1}}}\right).
\end{equation}

Now, by inserting scale factor (\ref{eqscal}) into the conditional
equation (\ref{eqfre}), one finds
\begin{equation}
c_{1}=\sqrt{\dfrac{2}{\lambda}},
\end{equation}
and therefore, scale factor can be written as
\begin{equation}
b(t)=\sqrt{\dfrac{2}{\lambda}}\left(\dfrac{e^{\sqrt{\frac{\lambda}{2}}(t+c_{2})}+\kappa e^{-\sqrt{\frac{\lambda}{2}}(t+c_{2})}}{2}\right),
\end{equation}
where $c_{2}$ is an integration constant and $ \kappa=0,\pm 1 $.

In the case of $\kappa=-1$ and $\lambda<0$, the scale factor
becomes
\begin{eqnarray}
b(t)=\sqrt{\dfrac{2}{\lambda}}\sin\left(\sqrt{\dfrac{2}{\lambda}}(t+c_{2})\right)
\end{eqnarray}
which is correspondence to the cyclic universe.\\
It is also notable that for $\kappa=0$ and $\lambda>0$, the
expansion of universe becomes an exponential form while in the
case of $\kappa=1$ and $\lambda>0$, the scale factor can be
simplified as
\begin{eqnarray}
b(t)=\sqrt{\dfrac{2}{\lambda}}\cosh\left(\sqrt{\dfrac{\lambda}{2}}(t+c_{2}) \right).
\end{eqnarray}

\section{Conclusion}

In this paper, we have presented five-dimensional thick brane
solutions supported by a scalar field. By choosing a special form
for the warp factor, we have obtained regular solutions with
finite energy density. These solutions are non-singular in the
whole space-time even at the location of the brane. Also, we have
investigated the stability of our thick brane solutions. The
effective potential of the gravitons shows that there are bound
states which are localized around the brane. Such gravitons make
the four-dimensional gravity on the brane Newtonian if we take the
thin brane and low energy limit. On the other hand, as $
z\rightarrow \infty $, the effective potential of the graviton
approaches zero, this means, there is no mass gap between the
excited KK modes and the massless ground state, while the
probability density of the KK states is a maximum at the brane
location. This means that the lighter KK excitations are closer to
the brane than the heavier ones and, hence, can interact with the
graviton with a greater probability. The interplay between the
probability of interaction and mass of the KK states is what
generates the effective mass gap. We have briefly
addressed the formalism of canonical gravity and the WD equation
as applied to the brane. We have seen that only in the case of $
\kappa=-1 $, tunneling occurs which means that the appropriate
classical cosmology subject to quantization is the spatially
spherical case.\\

Despite our warp function is not a linear function of extra
dimension when $w\rightarrow \infty$ (see eq. (\ref{eqapp12})),
the scalar potential is unbounded from below, i.e., it has not any
minimum. This means while the system rolls down, since there is
not a ground state for the system, it goes to negative infinity.
So, the scalar potential is not an appropriate potential
applicable to quantum effective field theory like Goldeston
potential \cite{Goldstone:1961eq}. Even though, in order to have a
good solution for thick brane, we should have following criteria:
having a warp function localized around the thick brane, kink like
scalar field and a scalar potential with at least two minima. It
is obvious that satisfaction of the above conditions,
simultaneously, is not possible for many candidates of warp
function \cite{Dzhunushaliev:2009va}.

It is interesting to calculate the amplitude of the bulk tensor
metric perturbations, the amplitude of the bound state modes and
tunneling rate. In the case of $\phi(w,t)=\phi(t)$ in Sec. 5, one
can go further by investigating the cosmological inflation of the
model in depth. We leave these works for future work.

\section*{Acknowledgements}
We are grateful to the anonymous referees for the insightful
comments. We also acknowledge fruitful discussion with K. Nozari.
SNS and NR thank the support of Shahid Beheshti University. SNS
and SHH thank Shiraz University Research Council.


\begin{thebibliography}{99}
\bibitem{1} T. Kaluza, Sitz. Preuss. Akad. Wiss. Phys. Math. {\bf K 1}. 966 (1921).
\bibitem{2} N. Arkani-Hamed, S. Dimopoulos and G. Dvali, Phys. Lett. {\bf B 429}, 263 (1998).
\bibitem{3} I. Antoniadis, S. Dimopoulos and G. Dvali, Nucl. Phys. {\bf B 516}, 70 (1998).
\bibitem{4} L. Randall and R. Sundrum, Phys. Rev. Lett. {\bf 83}, 3370 (1999).
\bibitem{5} L. Randall and R. Sundrum, Phys. Rev. Lett. {\bf 83}, 4690 (1999).
\bibitem{6} J. Garriga and T. Tanaka, Phys. Rev. Lett. {\bf 84}, 2778 (2000).
\bibitem{7} S. B. Giddings, E. Katz and L. Randall, JHEP {\bf 0003}, 023(2000).
\bibitem{8} O. DeWolfe, D. Z. Freedman, S. S. Gubser and A. Karch, Phys. Rev. {\bf D 62}, 046008 (2000).
\bibitem{9}  M. Gremm, Phys. Lett. {\bf B 478}, 434 (2000).
\bibitem{10} C. Csaki, J. Erlich, T. J. Hollowood and Y. Shirman, Nucl. Phys. {\bf B 581}, 309 (2000).
\bibitem{11} A. Kehagias and K. Tamvakis, Phys. Lett. {\bf B 504}, 38 (2001).
\bibitem{12} M. Gremm, Phys. Rev. {\bf D 62}, 044017 (2000).
\bibitem{24} V. Dzhunushaliev, Grav. Cosmol. {\bf 13}, 302 (2007).
\bibitem{25} V.~Dzhunushaliev, V.~Folomeev, S.~Myrzakul and R.~Myrzakulov,
  Mod.\ Phys.\ Lett.\ A {\bf 23}, 2811 (2008)
\bibitem{26} M. Gremm, Phys. Lett. {\bf B 478}, 434 (2000).
\bibitem{27} A. Wang, Phys. Rev. {\bf D 66}, 024024 (2002);
\bibitem{28} D. Bazeia, F. A. Brito and J.R. Nascimento, Phys. Rev. {\bf D 68}, 085007 (2003).
\bibitem{29} D. Bazeia, C. Furtado and A. R. Gomes, JCAP {\bf 0402}, 002 (2004).
\bibitem{30} S. Kobayashi, K. Koyama and J. Soda, Phys. Rev. {\bf D 65}, 064014 (2002).
\bibitem{31} A. Melfo, N. Pantoja and A. Skirzewski, Phys. Rev. {\bf D 67}, 105003 (2003).
\bibitem{32} C. Barcelo, C. Germani and C.F. Sopuerta, Phys. Rev. {\bf D 68}, 104007 (2003).
\bibitem{33} H. Guo, Y.-X. Liu, S.-W. Wei and C.-E. Fu, Europhys. Lett. {\bf 97}, 60003 (2012).
\bibitem{34} K.A. Bronnikov and B.E. Meierovich, Grav. Cosmol. {\bf 9}, 313 (2003).
\bibitem{35} A. Karch, L. Randall, JHEP, {\bf 0105}, 008 (2001).
\bibitem{36}  P. D. Mannheim, Brane-localized Gravity, (World Scientific Publishing Company, Singapore) 2005; SenGupta S., Aspects of warped brane world models, arXiv:0812.1092[hep-th].
\bibitem{13} J. A. Wheeler, Ann. Phys. (N.Y.) {\bf 2}, 604 (1957).
\bibitem{14} B. S. DeWitt, Phys. Rev. {\bf 160}, 1113 (1967).
\bibitem{15} J. B. Hartle, S. W. Hawking, Phys. Rev. {\bf D 28}, 2960 (1983).
\bibitem{16} S. W. Hawking, Nucl. Phys. {\bf B 239}, 257 (1984).
\bibitem{17} S. W. Hawking, D. N. Page, Nucl. Phys. {\bf B 264}, 185 (1986).
\bibitem{18} J. J. Halliwell, S. W. Hawking, Phys. Rev. {\bf D 31}, 1777 (1985).
\bibitem{19} A. Vilenkin, Phys. Lett. {\bf B 117}, 25 (1982).
\bibitem{20} A. Vilenkin, Phys. Rev. {\bf D 27}, 2848 (1983).
\bibitem{21} A. Vilenkin, Phys. Rev. {\bf D 30}, 549 (1984).
\bibitem{22} A. Vilenkin, Phys. Rev. {\bf D 37}, 888 (1988).
\bibitem{23} F. Darabi, W. N. Sajko and P. S. Wesson, Class. Quant. Grav. {\bf 17}, 4357 (2000).

\bibitem{Ahmed:2013lea}
  A.~Ahmed, B.~Grzadkowski and J.~Wudka,
  JHEP {\bf 1404}, 061 (2014).

  \bibitem{Peyravi:2015bra}
M.~Peyravi, N.~Riazi and F.~S.~N.~Lobo,
Eur. Phys. J. C \textbf{76}, no.5, 247 (2016)
doi:10.1140/epjc/s10052-016-4094-9
[arXiv:1504.04603 [gr-qc]].

\bibitem{Goldstone:1961eq}
J.~Goldstone,
Nuovo Cim. \textbf{19}, 154-164 (1961)
doi:10.1007/BF02812722

\bibitem{Dzhunushaliev:2009va}
V.~Dzhunushaliev, V.~Folomeev and M.~Minamitsuji,
Rept. Prog. Phys. \textbf{73}, 066901 (2010)
doi:10.1088/0034-4885/73/6/066901
[arXiv:0904.1775 [gr-qc]].

\end{thebibliography}
\end{document}